\documentclass[twocolumn,showpacs,preprintnumbers,amsmath,amssymb]{revtex4}

\usepackage{graphicx}
\usepackage{dcolumn}
\usepackage{bm}

\begin{document}
\newcommand{\gdhi}{\ooalign{\hfil/\hfil\crcr$\partial$}}

\def\Sp{\mathop{\mathrm{Sp}}\nolimits}
\def\sgn{\mathop{\mathrm{sgn}}\nolimits}
\def\erfc{\mathop{\mathrm{erfc}}\nolimits}
\def\tr{\mathop{\mathrm{tr}}\nolimits}
\def\as{\mathop{\mathrm{as}}\nolimits}
\def\val{\mathop{\mathrm{val}}\nolimits}

\title{Soliton solutions in an effective action for SU(2) Yang-Mills theory:
including effects of higher-derivative term}

\author{N.Sawado}
\email{sawado@ph.noda.tus.ac.jp}
\author{N.Shiiki}
\email{norikoshiiki@mail.goo.ne.jp}
\author{S.Tanaka}
\affiliation{
Department of Physics, Faculty of Science and Technology, 
Tokyo University of Science, Noda, Chiba 278-8510, Japan 
}
\date{\today}

\begin{abstract}
The Skyrme-Faddeev-Niemi (SFN) model which is an O(3) $\sigma$ model in three 
dimensional space upto fourth-order in the first derivative is regarded as a  
low-energy effective theory of SU(2) Yang-Mills theory. 
One can show from the Wilsonian renormalization group argument that 
the effective action of Yang-Mills theory recovers the SFN in the infrared 
region. However, the thoery contains an additional fourth-order term 
which destabilizes the soliton solution. 
In this paper, we derive the second derivative term perturbatively   
and show that the SFN model with the second derivative term 
possesses soliton solutions. 
\end{abstract}


\maketitle

\section{\label{sec:level1}Introduction\protect\\ } 
The Skyrme-Faddeev-Niemi (SFN) model which is an O(3) $\sigma$ model 
in three dimensional space upto fourth-order in the first derivative 
has topological soliton solutions with torus or knot-like structure. 
The model was initiated in 70's~\cite{faddeev75} and its interest has  
been extensively growing. The numerical simulations were performed in 
Refs.~\cite{faddeev97,gladikowski97,sutcliffe98,hietarinta99,hietarinta00}, 
the integrability was shown in Ref.~\cite{aratyn99}, and the application to 
the condensed matter physics~\cite{babaev02} and the Weinberg-Salam 
model~\cite{fayzullaev} were also considered.  
The recent research especially focuses on the consistency between 
the SFN and fundamental theories such as QCD~\cite{faddeev99,langmann99,
shabanov99,cho02}. 
In those references, it is claimed that the SFN action should be deduced  
from the SU(2) Yang-Mills (YM) action at low energies. 
One can also show from the Wilsonian renormalization group argument that 
the effective action of Yang-Mills theory recovers the SFN in the infrared 
region~\cite{gies01}. 
However, the derivative expansion for slowly varying 
fields $\bm{n}$ upto quartic order brings an additional fourth-order term 
in the SFN model to destabilize the soliton solution. 

Similar situations can be seen also in various topological soliton 
models. In the Skyrme model, the chirally invariant lagrangian 
with quarks produces fourth order terms after the derivative expansion 
and they destabilize the soliton solution~\cite{dhar85,aitchison85}.
To recover the stability of the skyrmion, the author of Ref.\cite{marleau01} 
introduced a large number of higher order terms in the first derivative 
whose coefficients were determined from the coefficients of the 
Skyrme model by using the recursion relations. 
Alternatively, in Ref.~\cite{gies01} Gies pointed out the possibility that 
the second derivative order term can work as a stabilizer for the soliton. 

In this paper, we examine the Gies's supposition by numerical analysis.
In section \ref{sec:level2}, we give an introduction to the Skyrme-Faddeev-Niemi 
model with its topological property. 
In section \ref{sec:level3}, we show how to derive the SFN model action 
from the SU(2) Yang-Mills theory. In section \ref{sec:level4}, 
soliton solutions of this truncated YM action are studied. 
In order to find stable soliton solutions, we introduce 
a second derivative term which can be derived in a perturbative 
manner.  
The naive extremization scheme, however, produce the fourth
order differential equation and the model has no stable soliton solution. 
Failure of finding the soliton is caused by the 
basic feature of the second derivative field theory. 
In section \ref{sec:level5}, the higher derivative theory and 
Ostrogradski's formulation are reviewed. We show the absence 
of bound state in the second derivative theory using an example in 
quantum mechanics and introduce the perturbative treatment for the 
second derivative theory. 
In section \ref{sec:level6}, we present our numerical results. 
In section \ref{sec:level7} are concluding remarks. 

\section{\label{sec:level2}Skyrme-Faddeev-Niemi model\protect\\}
The Faddeev-Niemi conjecture for the low-energy model of SU(2) 
Yang-Mills theory is expressed by following effective action:
\begin{eqnarray}
S_{\rm SFN}=\Lambda \int d^4x\Bigl[\frac{1}{2}(\partial_\mu \bm{n})^2
+\frac{g_1}{8}(\bm{n}\cdot \partial_\mu \bm{n}\times \partial_\nu \bm{n})^2 \Bigr] 
\label{fsn_ac}
\end{eqnarray}
where $\bm{n}(\bm{x})$ is a three component vector field normalized as 
$\bm{n}\cdot\bm{n}=1$. The mass scale $\Lambda$ is a free parameter 
and in this paper we set $\Lambda=1$. 
Stable soliton solutions exist when $g_1 > 0$. 

The static field $\bm{n}(\bm{x})$ maps $\bm{n}:R^3\mapsto S^2$ and 
the configurations are classified by the topological maps characterized 
by a topological invariant $H$ called Hopf charge
\begin{eqnarray}
H=\frac{1}{32\pi^2}\int A \wedge F,~~F=dA
\label{hopf}
\end{eqnarray}
where $F$ is the field strength and can be written as 
$F=(\bm{n}\cdot d\bm{n}\wedge d\bm{n})$.

The static energy $E_{\rm stt}$ from the action (\ref{fsn_ac}) has 
a topological lower bound~\cite{ward98}, 
\begin{eqnarray}
E_{\rm stt}\ge K H^{3/4}
\label{lowerbound}
\end{eqnarray}
where $K=16\pi^2\sqrt{g_1}$.

Performing numerical simulation, one can find that the static configurations 
for $H=1,2$ have axial symmetry~\cite{sutcliffe98}. 
Thus ``the toroidal ansatz'' which was studied in Ref.\cite{gladikowski97} 
is suitable to be imposed on these configurations. The ansatz is given by 
\begin{eqnarray}
&&n_1=\sqrt{1-w^2(\eta,\beta)}\cos(N\alpha+v(\eta,\beta))\,, \nonumber \\
&&n_2=\sqrt{1-w^2(\eta,\beta)}\sin(N\alpha+v(\eta,\beta)\,, 
\label{toroidal} \\
&&n_3=w(\eta,\beta)\,, \nonumber 
\end{eqnarray}
where $(\eta,\beta,\alpha)$ is toroidal coordinates which are related to 
the $R^3$ as follows:
\begin{eqnarray}
x=\frac{a\sinh\eta\cos\alpha}{\tau},y=\frac{a\sinh\eta\sin\alpha}{\tau},
z=\frac{a\sin\beta}{\tau}
\end{eqnarray}
with $\tau=\cosh\eta-\cos\beta$.

The function $w(\eta,\beta)$ is subject to the boundary conditions $w(0,\beta)=1,w(\infty,\beta)=-1$
and is periodic in $\beta$. $v(\eta,\beta)$ is set to be $v(\eta,\beta)=M\beta+v_0(\eta,\beta)$ and 
$v_0(,\beta)$ is considered as a constant map. 
Equation (\ref{hopf}) then gives $H=NM$.

In this paper we adopt a simpler ansatz than (\ref{toroidal}), which 
is defined by
\begin{eqnarray}
&&n_1=\sqrt{1-w^2(\eta)}\cos(N\alpha+M\beta)\,,\nonumber \\
&&n_2=\sqrt{1-w^2(\eta)}\sin(N\alpha+M\beta)\,, 
\label{afz} \\
&&n_3=w(\eta)\,, \nonumber 
\end{eqnarray}
where $w(\eta)$ satisfies the boundary conditions $w(0)=1,w(\infty)=-1$.
We numerically study soliton solutions for both ansatz (\ref{toroidal}) 
and (\ref{afz}). By comparing those results, we found that this simple 
ansatz produces at most 10\% errors and it does not affect to the 
property of the soliton solution. 

By using (\ref{afz}), the static energy is written in terms of the function 
$w(\eta)$ as
\begin{eqnarray}
&&E_{\rm stt}=2\pi^2a \int d\eta
\Biggl[
\frac{(w')^2}{1-w^2}+(1-w^2) U_{M,N}(\eta) \nonumber \\
&&\hspace{2cm}+\frac{g_1}{4a^2}\sinh\eta\cosh\eta (w')^2
U_{M,N}(\eta)\Biggr]\,,\nonumber \\
&&\hspace{1cm}w'\equiv \frac{dw}{d\eta},~~
U_{M,N}(\eta)\equiv \Bigl(M^2+\frac{N^2}{\sinh^2\eta}\Bigr)\,. \nonumber
\end{eqnarray}
The Euler-lagrange equation of motion is then derived as 
\begin{eqnarray}
&&\frac{w''}{1-w^2}+\frac{ww'^2}{(1-w^2)^2}+U_{M,N}(\eta)w \nonumber \\
&&+\frac{g_1}{2a^2}\Bigl(-2N^2\coth^2\eta w'+(\cosh^2\eta+\sinh^2\eta)
U_{M,N}(\eta)w' \nonumber \\ 
&&+\sinh\eta\cosh\eta U_{M,N}(\eta)w''\Bigr)=0\,.
\label{fsn_eq}
\end{eqnarray}
The variation with respect to $a$ produces the equation for variable $a$.
Soliton solutions are obtained by solving the equations for $a$ as 
well as for $w$.

\section{\label{sec:level3}Effective action in the Yang-Mills theory
with CFNS decomposition\protect\\}
In this section, we briefly review how to derive the
SFN effective action from the action of SU(2) Yang-Mills theory in the 
infrared limit~\cite{shabanov99,gies01}. 
For the gauge fields $\bm{A}_\mu$, the Cho-Faddeev-Niemi-Shabanov 
decomposition is applied~\cite{faddeev99,langmann99,shabanov99,cho02}
\begin{eqnarray}
\bm{A}_\mu=\bm{n}C_\mu+(\partial_\mu\bm{n})\times\bm{n}+\bm{W}_\mu\,.
\label{cfns}
\end{eqnarray}
The first two terms are the ``electric'' and ``magnetic'' Abelian connection, 
and $\bm{W}_\mu$ are chosen so as to orthogonal to $\bm{n}$, 
$\bm{W}_\mu\cdot\bm{n}=0$.
Obviously, the degrees of freedom on the left- and right-hand side 
of Eq.(\ref{cfns}) do not match. While the LHS describes 
$3_{\rm color}\times 4_{\rm Lorentz}=12$, the RHS is comprised of 
$(C_\mu:)4_{\rm Lorentz}+(\bm{n}:)2_{\rm color}+(\bm{W}_\mu:)3_{\rm color}
\times4_{\rm Lorentz}-4_{\bm{n}\cdot\bm{W}_\mu=0}=14$ degrees freedom. 
Shabanov introduced in his paper \cite{shabanov99} the following constraint 
\begin{eqnarray}
\bm{\chi}(\bm{n},C_\mu,\bm{W}_\mu)=0,~{\rm with}~~\bm{\chi}\cdot\bm{n}=0\,.
\end{eqnarray}
The generating functional of YM theory can be written by using 
Eq.(\ref{cfns}) as 
\begin{eqnarray}
{\cal Z}=\int {\cal D}\bm{n}{\cal D}C{\cal D}\bm{W}\delta(\bm{\chi})
\Delta_{\rm FP}\Delta_{\rm S}e^{-S_{\rm YM}-S_{\rm gf}}\,.
\label{vf0}
\end{eqnarray}
$\Delta_{\rm FP}$ and $S_{\rm gf}$ are the Faddeev-Popov determinant and 
the gauge fixing action respectively, and Shabanov introduced another 
determinant $\Delta_{\rm S}$ corresponding to the condition $\bm{\chi}=0$.
YM and the gauge fixing action is given by 
\begin{eqnarray}
&&S_{\rm YM}+S_{\rm  gf}=\int d^4x\Bigl[\frac{1}{4g^2}\bm{F}_{\mu\nu}\cdot\bm{F}_{\mu\nu}
+\frac{1}{2\alpha_{\rm g} g^2}(\partial_\mu \bm{A}_\mu)^2\Bigr] \,. \nonumber 
\end{eqnarray}
Inserting Eq.(\ref{cfns}) into the action, one obtains the vacuum functional
\begin{eqnarray}
&&{\cal Z}=\int {\cal D}\bm{n}e^{-{\cal S}_{\rm eff}(\bm{n})} \nonumber \\
&&~~~=\int {\cal D}\bm{n} e^{-{\cal S}_{\rm cl}(\bm{n})}\int {\cal D}\tilde{C}{\cal D}\bm{W}_\mu
\Delta_{\rm FP}\Delta_{\rm S}\delta({\bm \chi}) \nonumber \\
&&~~~\times e^{-(1/2g^2)\int(\tilde{C}_\mu M^C_{\mu\nu}\tilde{C}_\nu+\bm{W}_\mu \bar{M}^{\bm{W}}_{\mu\nu}\bm{W}_\nu
+2C_\nu K^C_\nu+2\bm{W}_\mu\cdot\bm{K}^{\bm{W}}_\mu)}
\nonumber \\
\end{eqnarray}
with
\begin{eqnarray}
&&M^C_{\mu\nu}=-\partial^2\delta_{\mu\nu}+\partial_\mu\bm{n}\cdot\partial_\nu\bm{n}\,, \nonumber \\
&&M^{\bm{W}}_{\mu\nu}=-\partial^2\delta_{\mu\nu}-\partial_\mu\bm{n}\otimes\partial_\nu\bm{n}
+\partial_\nu\bm{n}\otimes\partial_\mu\bm{n}\,, \nonumber \\
&&\bm{Q}^C_{\mu\nu}=\partial_\mu\bm{n}\partial_\nu+\partial_\nu\bm{n}\partial_\mu+\partial_\mu\partial_\nu\bm{n}\,,
\\
&&K^C_{\mu\nu}=\partial_\nu(\bm{n}\cdot\partial_\nu\bm{n}\times\partial_\mu\bm{n})
+\partial_\mu\bm{n}\cdot\partial^2\bm{n}\times\bm{n}\,, \nonumber \\
&&\bm{K}^{\bm W}_{\mu\nu}=\partial_\mu(\bm{n}\times\partial^2\bm{n})\,, 
\hspace{1cm}{\rm (in~gauge~\alpha_g=1}) \nonumber 
\end{eqnarray}
and 
\begin{eqnarray}
&&\bar{M}^{\bm{W}}_{\mu\nu}:=M^{\bm{W}}_{\mu\nu}-\bm{Q}_{\mu s}{M^C}^{-1}_{s\lambda}\bm{Q}_{\lambda \nu}\,, \nonumber \\
&&\tilde{C}_\mu=C_\mu+\bm{W}_s\cdot \bm{Q}_{s\lambda}{M^C}^{-1}_{\lambda\mu}\,.
\end{eqnarray}
The classical action of $\bm{n}$ including the gauge fixing term is given by
\begin{eqnarray}
{\cal S}_{\rm cl}=\int d^4x\Bigl[\frac{1}{4g^2}(\partial_\mu\bm{n}\times \partial_\nu\bm{n})^2
+\frac{1}{2 \alpha_{\rm g} g^2}(\partial^2\bm{n}\times\bm{n})^2\Bigr]\,.\nonumber \\
\end{eqnarray}
The $\delta$ functional is expressed by its Fourier transform
\begin{eqnarray}
\delta(\bm{\chi})=\int {\cal D}\bm{\phi}e^{-i\int (\bm{\phi}\cdot \partial\bm{W}_\mu
+\bm{\phi}\cdot C_\mu\bm{n}\times \bm{W}_\mu+(\bm{\phi}\cdot\bm{n})(\partial_\mu\bm{n}\cdot\bm{W}_\mu))}\,.
\nonumber \\
\end{eqnarray} 
Integrating over $C,\bm{W},\bm{\phi}$, we finally obtain 
\begin{eqnarray}
&&e^{-S_{\rm eff}}=e^{-S_{\rm cl}}\Delta_{\rm FP}\Delta_{\rm S}
(\det M^C)^{-1/2} (\det \bar{M}^{\bm{W}})^{-1/2}
\nonumber \\
&&\hspace{1cm}\times (\det -Q^{\bm{\phi}}_\mu (\bar{M}^{\bm{W}})^{-1}_{\mu\nu}
Q^{\bm{\phi}}_\nu)^{-1/2} \label{determ} 
\end{eqnarray}
where several nonlocal terms and the higher derivative components have been 
neglected. 

We perform the derivative expansion for the four determinants in 
Eq.~(\ref{determ}) under the following assumptions 
\begin{itemize}
\item [(i)]the theory is valid for the momenta $p$ with $k<p<\Lambda$ ($k,\Lambda$ are 
infrared and ultraviolet cut-off)
\item [(ii)]$|\partial\bm{n}| \ll k$ 
\item [(iii)]the higher derivative terms, such as $\partial^2\bm{n}$ are omitted.
\end{itemize}
The effective action is then given by  
\begin{eqnarray}
&&S_{\rm eff}=\int d^4x\Bigl[\frac{1}{2}(\partial_\mu \bm{n})^2
+\frac{g_1}{8}(\partial_\mu \bm{n}\times \partial_\nu \bm{n})^2 \nonumber \\
&&\hspace{2cm}+\frac{g_2}{8}(\partial_\mu \bm{n})^4
 \Bigr]\,.
\label{fsn2}
\end{eqnarray}
For $g_1>0$ and $g_2=0$, the action is identical to the FSN effective 
action~(\ref{fsn_ac}). 

In order to get the stable soliton solutions, $g_2$ must be 
positive~\cite{gladikowski97}. However, $g_2$ is found to be 
negative according to the above analysis.  
Therefore we consider higher-derivative terms and investigate 
if the model with the higher-derivatives possess soliton solutions. 

\section{\label{sec:level4}Search for the stable soliton solutions (1)\protect\\}
The static energy is derived from Eq.(\ref{fsn2}) as  
\begin{eqnarray}
E_{\rm stt}&=& \int d^3x\Bigl[\frac{1}{2}(\partial_i \bm{n})^2
+\frac{g_1}{8}(\partial_i \bm{n}\times \partial_j \bm{n})^2
+\frac{g_2}{8}(\partial_i \bm{n})^4 \Bigr] \nonumber \\
&:=&E_2(\bm{n})+E_4^{(1)}(\bm{n})+E_4^{(2)}(\bm{n})\,.
\end{eqnarray}
A spatial scaling behaviour of the static energy, so called Derrick's scaling 
argument, can be applied to examine the stability of the soliton~\cite{sutcliffe05}. 
Considering the map 
$\bm{x}\mapsto \bm{x}'=\mu\bm{x}~(\mu>0)$, with 
$\bm{n}^{(\mu)}\equiv \bm{n}(\mu\bm{x})$, the static energy scales as 
\begin{eqnarray}
e(\mu)&=&E_{\rm stt}(\bm{n}^{(\mu)}) \nonumber \\
&=&E_2(\bm{n}^{(\mu)})+E_4^{(1)}(\bm{n}^{(\mu)})+E_4^{(2)}(\bm{n}^{(\mu)}) \nonumber \\
&=&\frac{1}{\mu}E_2(\bm{n})+\mu(E_4^{(1)}(\bm{n})+E_4^{(2)}(\bm{n}))\,.
\label{derrick}
\end{eqnarray}
Derrick's theorem states that if the function $e(\mu)$ has no stationary point, 
the theory has no static solutions of the field equation with finite density, 
other than the vacuum. 
Conversely, if $e(\mu)$ has stationary point, the  possibility of 
having finite energy soliton solutions is not excluded. 
Eq.(\ref{derrick}) is stationary at $\mu=\sqrt{E_2/(E_4^{(1)}+E_4^{(2)})}$. 
Then, the following inequality
\begin{eqnarray}
&&g_1(\partial_i\bm{n}\times\partial_j\bm{n})^2+g_2(\partial_i\bm{n})^2(\partial_j\bm{n})^2 \nonumber \\
&&=g_1(\partial_i\bm{n})^2(\partial_j\bm{n})^2-g_1(\partial_i\bm{n}\cdot\partial_j\bm{n})^2
+g_2(\partial_i\bm{n})^2(\partial_j\bm{n})^2 \nonumber \\
&&\geqq g_2(\partial_i\bm{n}\cdot\partial_j\bm{n})^2
~~~~(\because (\partial_i\bm{n})^2(\partial_j\bm{n})^2\geqq (\partial_i\bm{n}\cdot\partial_j\bm{n})^2) \nonumber \\
\end{eqnarray}
ensures the possibility of existence of the stable soliton solutions for $g_2\geqq 0$. 
As mentioned in the section \ref{sec:level3}, $g_2$ should be negative at least within 
our derivative expansion analysis of YM theory. 

A promising idea to tackle the problem was suggested by Gies~\cite{gies01}.
He considered the following type of effective action, accompanying second 
derivative term 
\begin{eqnarray}
&&S_{\rm eff}=\int d^4x\Bigl[\frac{1}{2}(\partial_\mu \bm{n})^2
+\frac{g_1}{8}(\partial_\mu \bm{n}\times \partial_\nu \bm{n})^2 
\nonumber \\
&&\hspace{2cm}-\frac{g_2}{8}(\partial_\mu \bm{n})^4
+\frac{g_2}{8}(\partial^2 \bm{n}\cdot\partial^2 \bm{n})
 \Bigr]\,.
\label{fsn_ac2}
\end{eqnarray}
Here we choose positive value of $g_2$ and assign the explicit negative sign 
to the third term. In principle, it is possible to estimate the second derivative term 
by the derivative expansion without neglecting 
throughout the calculation.
The calculation is, however, very laborious and hence we show only 
one simple example of the $C$ determinant.
The determinant is real and thus it is expanded as follows
\begin{eqnarray}
&&\log(\det M^C)^{-1/2}
=-\frac{1}{2}{\rm Tr}\log(-\partial^2+\partial_\mu\bm{n}\cdot\partial_\mu\bm{n}) \nonumber \\ 
&&~~\to-\frac{1}{4}{\rm Tr}\log[\partial^4
-2(\partial\bm{n})^2\partial^2+(\partial\bm{n})^4-\partial^2(\partial\bm{n})^2]
\nonumber \\
&&~~=-\frac{1}{4}{\rm Tr}\log(\partial^4) \nonumber \\
&&~~~~~-\frac{1}{4}{\rm Tr}\log\Bigl[1-2\frac{(\partial\bm{n}^2)}{\partial^2}
+\frac{(\partial\bm{n})^4}{\partial^4}
-\frac{\partial^2(\partial\bm{n})^2}{\partial^4}\Bigr] \nonumber \\
&&~~=-\frac{1}{4}{\rm Tr}\log(\partial^4) \nonumber \\
&&~~~~~-\frac{1}{4}{\rm Tr}\Bigl[-2\frac{(\partial\bm{n}^2)}{\partial^2}
+\frac{(\partial\bm{n})^4}{\partial^4}
-\frac{\partial^2(\partial\bm{n})^2}{\partial^4}\Bigr] \nonumber \\
&&~~~~~+\frac{1}{8}{\rm Tr}\Bigl[-2\frac{(\partial\bm{n})^2}{\partial^2}\Bigr]^2
+O((\partial\bm{n})^6)
\end{eqnarray}
where we have defined $\partial_\mu\bm{n}\cdot\partial_\mu\bm{n}\to(\partial\bm{n})^2,
(\partial_\mu\bm{n}\cdot\partial_\nu\bm{n})^2\to(\partial\bm{n})^4$.
Employing the integral formulas~\cite{gies01}
\begin{eqnarray}
&&\int_{[k,\Lambda]} \frac{d^4p}{(2\pi)^4}\frac{1}{p^2}=\frac{1}{16\pi^2}(\Lambda^2-k^2)\,, \nonumber \\
&&\int_{[k,\Lambda]} \frac{d^4p}{(2\pi)^4}\frac{1}{p^4}=\frac{1}{8\pi^2}\log\frac{\Lambda}{k} \nonumber 
\end{eqnarray}
together with the equality (up to second derivative)
\begin{eqnarray}
\partial^2(\partial_\mu\bm{n}\cdot\partial_\mu\bm{n})=-\partial^2(\bm{n}\cdot\partial^2\bm{n})
= -\partial^2\bm{n}\cdot\partial^2\bm{n}\,,
\end{eqnarray}
we obtain the form for the $C$ determinant
\begin{eqnarray}
&&\log(\det M^C)^{-1/2}= \nonumber \\
&&-\frac{1}{32\pi^2}\int d^4x\Bigl[(\Lambda^2-k^2) (\partial_\mu\bm{n})^2
-\log\frac{\Lambda}{k}(\partial_\mu\bm{n}\times\partial_\nu\bm{n})^2  \nonumber \\ 
&&\hspace{1.5cm}+\log\frac{\Lambda}{k}(\partial_\mu\bm{n})^4
-\log\frac{\Lambda}{k}(\partial^2\bm{n}\cdot\partial^2\bm{n})\Bigr]\,.
\end{eqnarray}
The other determinants can be estimated in a similar manner. 

The static energy of Eq.(\ref{fsn_ac2}) with the ansatz (\ref{afz}) is written as 
\begin{eqnarray}
&&E_{\rm stt}=2\pi^2a \int d\eta
\Biggl[
\frac{(w')^2}{1-w^2}+(1-w^2) U_{M,N}(\eta) \nonumber \\
&&+\frac{g_1}{4a^2}\sinh\eta\cosh\eta (w')^2
U_{M,N}(\eta)\nonumber \\
&&-\frac{g_2}{4a^2}\sinh\eta\cosh\eta\biggl[\frac{(w')^2}{1-w^2}
+(1-w^2)U_{M,N}(\eta)\biggr]^2 \nonumber \\
&&+\frac{g_2}{4a^2}\biggl[\Bigl(\coth\eta+\sinh^2\eta-\sinh\eta\cosh\eta\Bigr)
\frac{(w')^2}{1-w^2} \nonumber \\
&&\hspace{1cm}+(\sinh\eta\cosh\eta-\sinh^2\eta)(1-w^2)M^2 \nonumber \\
&&\hspace{1cm}+2\Bigl\{ \frac{w(w')^3}{(1-w^2)^2}
+\frac{w' w''}{1-w^2}
+w w'U_{M,N}(\eta) \Bigr\}\nonumber \\
&&\hspace{1cm}+\sinh\eta\cosh\eta\Bigl\{ \frac{1}{1-w^2}\Bigl[\frac{(w')^2}{1-w^2}
+w w'' \nonumber \\
&&\hspace{1.5cm}+(1-w^2)U_{M,N}(\eta)\Bigr]^2 
+(w'')^2\Bigr\}\biggr]
\Biggl], \nonumber \\
&&\hspace{4cm}w''\equiv \frac{d^2w}{d\eta^2}.\nonumber 
\nonumber 
\end{eqnarray} 
The Euler-Lagrange equation of motion is derived by 
\begin{eqnarray}
-\frac{d^2}{d\eta^2}\Bigl(\frac{\partial E_{\rm stt}}{\partial w''}\Bigr)
+\frac{d}{d\eta}\Bigl(\frac{\partial E_{\rm stt}}{\partial w'}\Bigr)
-\frac{\partial E_{\rm stt}}{\partial w}=0 \,,
\end{eqnarray}
which is too complicated and thus we adopt the following notation  
\begin{eqnarray}
&&f_0(w,w',w'')+g_1f_1(w,w',w'') \nonumber \\
&&\hspace{1.5cm}+g_2f_2(w,w',w'',w^{(3)},w^{(4)})=0\,.
\label{fsn_eq2}
\end{eqnarray}
Here $w^{(3)},w^{(4)}$ represent the third and the fourth derivative with 
respect to $\eta$. The first two terms of Eq.(\ref{fsn_eq2}) are  
identical to those in Eq.(\ref{fsn_eq}).

Unfortunately, we could not find out stable soliton solutions from 
Eq.(\ref{fsn_eq2}) for any value of $g_2$. 

From the relation
\begin{eqnarray}
\int d^4x[(\partial^2 \bm{n}\cdot\partial^2 \bm{n})-(\partial_\mu \bm{n})^4]
=\int d^4x(\partial^2 \bm{n}\times\bm{n})^2\,,
\end{eqnarray} 
one easily finds that the static energy obtained from the last two terms 
in Eq.(\ref{fsn_ac2}) 
\begin{eqnarray}
\tilde{E}^{(2)}_4=\int d^3x(\partial^2 \bm{n}\times\bm{n})^2
\label{energy2}
\end{eqnarray}
gives the positive contribution. The total static energy is stationary 
at $\mu=\sqrt{E_2/(E_4^{(1)}+\tilde{E}_4^{(2)})}$ and hence the possibility 
of existence of soliton solutions is not excluded. 
And also, the positivity of Eq.(\ref{energy2}) does not spoil the lower 
bound (\ref{lowerbound}) of original SFN and the possibility is still  
not excluded, too.  

Therefore, we suspect that the absence of the stable soliton is caused 
by the fact that {\it higher derivative theory has no lower bound state}. 
We shall investigate the lower bound in the higher derivative theory 
in detail in the next section. 

\section{\label{sec:level5}Higher derivative theory\protect\\}
In this section, we address the basic problems in the higher derivative 
theory~\cite{pais,smilga,eliezer89,jaen,simon} which essentially falls 
into two categories. The first problem concerns the increase in the  
number of degrees of freedom. For example, if the theory contains 
second derivative terms, the equation of motion becomes up to the order 
in the fourth derivative. Thus, four parameters are required for 
the initial conditions. If one considers more higher order terms, 
the situation gets worse. However, this is not serious problem for 
our study because our concern is the existence of static soliton solutions. 
The second problem is that the actions of the theory are not bounded from 
below. This feature makes the higher derivative theories unstable.

The lagrangian and the hamiltonian formalism with higher derivative was firstly developed
by Ostrogradski~\cite{ostrogradski}. We consider the lagrangian containing up to $n$th order derivatives
\begin{eqnarray}
S=\int dt {\cal L}(q,\dot{q},\cdots,q^{(n)})\,.
\end{eqnarray} 
Taking the variation of the action $\delta S=0$ leads the Euler-lagrange equation of motion
\begin{eqnarray}
\sum_{i=0}^n (-1)^i\frac{d^i}{dt^i}\Bigl(\frac{\partial {\cal L}}{\partial q^{(i)}}\Bigr)=0\,.
\end{eqnarray}
The hamiltonian is obtained by introducing $n$ generalized momenta
\begin{eqnarray}
p_i=\sum_{j=i+1}^n (-1)^{j-i-1}\frac{d^{j-i-1}}{dt^{j-i-1}}\Bigl(\frac{\partial {\cal L}}{\partial q^{(j)}}\Bigr)\,,
i=1,\cdots,n,
\end{eqnarray}
or
\begin{eqnarray}
&&p_n=\frac{\partial {\cal L}}{\partial q^{(n)}}\,, \nonumber \\
&&p_i=\frac{\partial {\cal L}}{\partial q^{(i)}}-\frac{d}{dt}p_{i+1}\,,~~i=1,\cdots,n-1,
\label{canonical momenta}
\end{eqnarray}
and $n$ independent variables
\begin{eqnarray}
&&q_1\equiv q\,,\nonumber \\
&&q_i\equiv q^{(i-1)}\,,~~(i=2,\cdots,n)\,.
\end{eqnarray}
The lagrangian now depends on the $n$ coordinates $q_i$ and on the first derivative $\dot{q}_n=q^{(n)}$.
The hamiltonian is defined as
\begin{eqnarray}
{\cal H}(q_i,p_i)=\sum^n_{i=1}p_i\dot{q}_i-{\cal L}=\sum^{n-1}_{i=1} p_i q_{i+1}+p_n \dot{q}_n-{\cal L}\,.
\end{eqnarray}
The canonical equations of motion turn out to be
\begin{eqnarray}
\dot{q}_i=\frac{\partial {\cal H}}{\partial p_i}\,,~~\dot{p}_i=-\frac{\partial {\cal H}}{\partial q_i}\,.
\end{eqnarray}

We consider a simple example including second derivative term \cite{simon}, defined as 
\begin{eqnarray}
{\cal L}=\frac{1}{2}(1+\varepsilon^2\omega^2)\dot{q}^2-\frac{1}{2}\omega^2q^2-\frac{1}{2}\varepsilon^2\ddot{q}^2\,,
\end{eqnarray}
where constant $\epsilon$ works as a coupling constant of second derivative term. 
The equation of motion is
\begin{eqnarray}
(1+\varepsilon^2 \omega^2)\ddot{q}+\omega^2 q+\varepsilon^2 q^{(4)}=0\,.
\label{eq_quanta}
\end{eqnarray}
From Eq.~(\ref{canonical momenta}), one gets 
\begin{eqnarray}
&&\pi_{\dot{q}}=\frac{\partial{\cal L}}{\partial \ddot{q}}=-\varepsilon^2\ddot{q}\,, \nonumber \\
&&\pi_q=\frac{\partial{\cal L}}{\partial \dot{q}}-\frac{d}{dt}\Bigl(\frac{\partial{\cal L}}{\partial \ddot{q}}\Bigr)
=(1+\varepsilon^2\omega^2)\dot{q}+\varepsilon^2\dddot{q}\,.
\end{eqnarray}
Thus the hamiltonian becomes  
\begin{eqnarray}
{\cal H}&=&\dot{x}\pi_q+\ddot{q}\pi_{\dot{q}}-{\cal L} \nonumber \\
&=&\dot{q}\pi_q-\frac{1}{2\varepsilon^2}\pi_{\dot{q}}^2
-\frac{1}{2}(1+\varepsilon^2\omega^2)\dot{q}^2+\frac{1}{2}\omega^2q^2\,.
\end{eqnarray}
We introduce the new canonical variables
\begin{eqnarray}
&&q_+=\frac{1}{\omega\sqrt{1-\varepsilon^2\omega^2}}(\varepsilon^2\omega^2\dot{q}-\pi_q)~,\nonumber \\
&&p_+=\frac{w}{\sqrt{1-\varepsilon^2\omega^2}}(q-\pi_{\dot{q}})\,, \nonumber \\
&&q_-=\frac{\varepsilon}{\sqrt{1-\varepsilon^2\omega^2}}(\dot{q}-\pi_q)~, \nonumber \\
&&p_-=\frac{1}{\varepsilon\sqrt{1-\varepsilon^2\omega^2}}(\varepsilon^2\omega^2 q-\pi_{\dot{q}})\,,
\nonumber 
\end{eqnarray}
and the hamiltonian has of the form by using these variables
\begin{eqnarray}
{\cal H}\to \frac{1}{2}(p_+^2+\omega^2 q_+^2)-\frac{1}{2}(p_-^2+\frac{1}{\varepsilon^2} q_-^2)\,. \nonumber 
\end{eqnarray}
The corresponding energy spectra is then given by
\begin{eqnarray}
E=(n+\frac{1}{2})\omega-(m+\frac{1}{2})\frac{1}{\varepsilon}~,~~n,m=0,1,2,\cdots
\end{eqnarray}
One can see that in the limit $\epsilon\to 0$ the energy goes to 
negative infinity rather than approaching to the harmonic oscillator 
energy eigenstates.  

To obtain physically meaningful solutions, we employ the perturbative 
analysis where the solution is expanded in terms of the small coupling 
constant and the Euler-Lagrange equation of motion is replaced with the 
corresponding perturbative equation. 
The solutions of the equations of motion that are ill behaved in the limit 
$\epsilon\to 0$ are excluded from the very beginning~\cite{eliezer89,jaen,simon}.

We assume that the solution of Eq.(\ref{eq_quanta}) can be written as 
\begin{eqnarray}
q_{\rm pert}(t)=\sum^{\infty}_{n=0} \epsilon^n q(t)\,. \label{q}
\end{eqnarray}
Substituting Eq.(\ref{q}) into Eq.(\ref{eq_quanta}) and taking time 
derivatives of these equations, we obtain the constraints for higher 
derivative terms
\begin{eqnarray}
&&O(\epsilon^0)  \nonumber \\
&&\hspace{3mm}equation :~~\ddot{q}_0+\omega^2q_0=0\,, \label{cnst00} \\
&&\hspace{3mm}constraints :~~\dddot{q}_0=-\omega^2\dot{q}_0, \ddddot{q}_0=\omega^4 q_0\,. \label{cnst01} \\
&&O(\epsilon^2)  \nonumber \\
&&\hspace{3mm}equation :~~\ddot{q}_2+\omega^2\ddot{q}_0+\omega^2 q_2+\ddddot{q}_0=0\,,  \nonumber \\
&&\hspace{13mm}\Rightarrow \ddot{q}_2+\omega^2q_2=0\,,~~({\rm using}~(\ref{cnst00}),(\ref{cnst01}))\,, \label{cnst20} \\
&&\hspace{3mm}constraints :~~\dddot{q}_2=-\omega^2\dot{q}_2, \ddddot{q}_2=\omega^4 q_2\,. \label{cnst21} \\
&&O(\epsilon^4)  \nonumber \\
&&\hspace{3mm}equation :~~\ddot{q}_4+\omega^2\ddot{q}_2+\omega^2 q_4+\ddddot{q}_2=0\,,  \nonumber \\
&&\hspace{13mm}\Rightarrow \ddot{q}_4+\omega^2q_4=0\,,~~({\rm using}~(\ref{cnst20}),(\ref{cnst21}))\,, \\
&&\hspace{3mm}constraints :~~\dddot{q}_4=-\omega^2\dot{q}_4, \ddddot{q}_4=\omega^4 q_4\,. \label{cnst4}
\end{eqnarray}
Combining these results, we find the perturbative equation of motion up to $O(\epsilon^4)$
\begin{eqnarray}
\ddot{q}_{\rm pert}+\omega^2q_{\rm pert}=O(\epsilon^6)\,.
\end{eqnarray}
which is the equation for harmonic oscillator. 

\begin{figure}
\includegraphics[height=7cm, width=9cm]{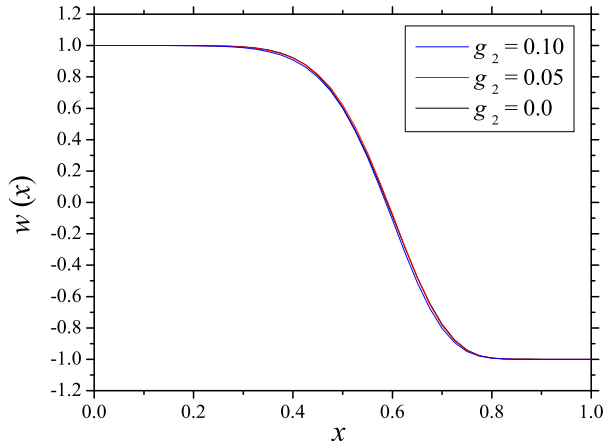}
\caption{\label{fig:Fig1} The function $w(\eta)$ for $g_1=0.4, g_2=0,0.05,0.1$
(the rescaling radial coordinate $x=\eta/(1-\eta)$ is used). }
\end{figure}
\begin{figure}
\includegraphics[height=7cm, width=9cm]{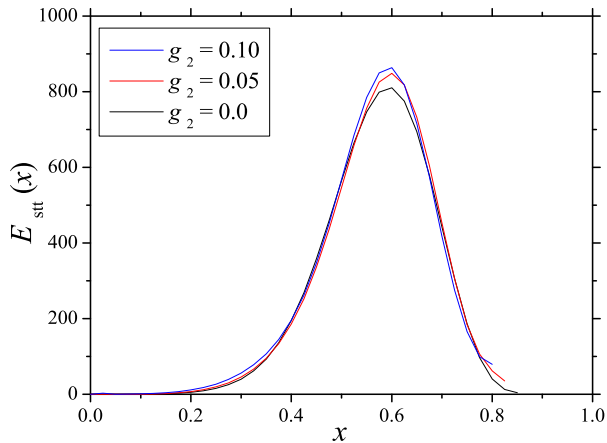}
\caption{\label{fig:Fig2} The energy density for $g_1=0.4, g_2=0,0.05,0.1$.}
\end{figure}

\begin{figure}
\includegraphics[height=7cm, width=9cm]{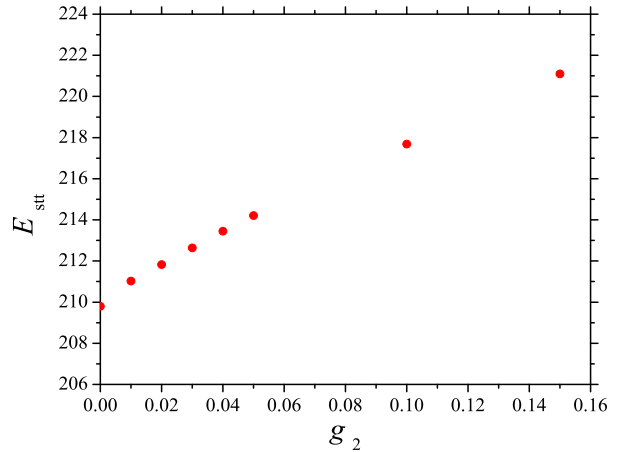}
\caption{\label{fig:Fig3} The energy as a function of $g_2$($g_1=0.4$).}
\end{figure}

\section{\label{sec:level6}Search for the stable soliton solutions (2) -- perturbative analysis --\protect\\}
Let us employ the perturbative method introduced in the last section to our 
problem. We assume that $g_2$ is relatively small and can be considered as 
a perturbative coupling constant. Thus, the perturbative solution is written 
by a power series in $g_2$
\begin{eqnarray}
w(\eta)=\sum^\infty_{n=0}g_2^n w_n(\eta)\,.
\label{sol_expand}
\end{eqnarray}
Substituting Eq.(\ref{sol_expand}) into Eq.(\ref{fsn_eq2}), we obtain 
the classical field equation in $O(g^0_2)$
\begin{eqnarray}
f_0(w_0,w_0',w_0'')+g_1f_1(w_0,w_0',w_0'')=0\,. \label{classical}
\end{eqnarray}
Taking derivatives for both sides in Eq.(\ref{classical}) and solving 
for $w_0'',w_0^{(3)},w_0^{(4)}$ read the following form
of the constraint equations for higher derivatives
\begin{eqnarray}
w_0^{(i)}=F^{(i)}(w_0,w_0')\,,~~i=2,3,4\,.
\label{constraint}
\end{eqnarray}
The equation in $O(g^1_2)$ can be written as 
\begin{eqnarray}
(f_0+g_2 f_1)_{O(g^1_2)}+f_2(w_0,w_0',w_0'',w_0^{(3)},w_0^{(4)})=0\,.
\label{fsn_eq21}
\end{eqnarray}
Substituting the constraint equations (\ref{constraint}) into Eq.(\ref{fsn_eq21})
and eliminate the higher derivative terms, one can obtain the perturbative 
equation of motion
\begin{eqnarray}
f_0(w,w',w'')+g_1 f_1(w,w',w'')+g_2 \tilde{f}_2(w,w')=O(g^2_2)\,.\nonumber \\
\label{fsn_eq2_p}
\end{eqnarray}

Now Eq.(\ref{fsn_eq2_p}) has topological soliton solutions. 
Our results of the estimated function $w(\eta)$ and the energy density are displayed 
in Figs.\ref{fig:Fig1},\ref{fig:Fig2}.(In all figures, we show the results for the 
case of Hopf charge $H=2;N=2,M=1$). We have small changes for varying the 
coupling constant $g_2$. The dependence of the $g_2$ for the total energy 
is shown in Fig.\ref{fig:Fig3}. 
It can be seen that the change is moderate with respect to $g_2$.

\section{\label{sec:level7}Summary\protect\\}
In this paper we have studied the Skyrme-Faddeev-Niemi
model and its extensions by introducing the reduction scheme of the SU(2) 
Yang-Mills theory to the corresponding low-energy effective model. 
The requirement of consistency between the low-energy effecive action 
of the YM and the SFN type model lead us to take into account second 
derivative terms in the action. 
However, we found that such an action including the second derivative terms 
does not have stable soliton solutions.
This is due to the absence of the energy bound in higher 
derivative theory. 
This fact inspired us to employ the perturbative analysis to our effective 
action. Within the perturbative analysis, we were able to obtain the topological 
soliton solutions. 
 
Our analysis is based on perturbation and the coupling constant $g_2$ is 
assumed to be small.  
However, Wilsonian renormalization analysis of YM theory~\cite{gies01} 
suggest that the coupling constants $g_1,g_2$ (and the mass scale parameter 
$\Lambda$) depend on the renormalization group time $t=\log k/\Lambda$ 
($k,\Lambda$ are infrared, ultraviolet cutoff parameter) and those are 
almost comparable. To improve the analysis, we could perform the 
next order of perturbation, but it is tedius and spoils the simplicity 
of the FSN model unfortunately. 

It should be noted that our solutions do not much differ from 
the solution of original SFN model, at least in the perturbative regime. 
We suspect that some appropriate truncation ({\it like} ``extra fourth order term 
+ second derivative term'') always supply the stable solutions that are 
close to the original SFN model. Thus we conclude that the topological 
soliton model comprised of the ``kinetic term + a special fourth order 
term'' like SFN model is a good approximation. 

\begin{acknowledgments}
The authors thank to Kei-Ichi Kondo for drawing our attention to the 
coefficient problem of this model.
\end{acknowledgments}

\end{document}